\documentclass[
aps,%
12pt,%
final,%
notitlepage,%
oneside,%
onecolumn,%
nobibnotes,%
nofootinbib,%
superscriptaddress,%
noshowpacs,%
centertags]{revtex4-2}
\usepackage{braket}
\usepackage{graphicx}
\usepackage{subcaption}
\usepackage{amssymb,amsthm,amsmath,bm}
\begin{document}

\title{The influence of majorons on neutrino oscillations in the presence of a magnetic field and matter}

\author{\firstname{A.~A.}~\surname{Lichkunov}}
\email{lichkunov.aa15@physics.msu.ru}
\affiliation{Department of Theoretical Physics, Moscow State University, 119991 Moscow, Russia}%
\author{\firstname{A.~R.}~\surname{Popov}}
\email{ar.popov@physics.msu.ru}
\affiliation{Department of Theoretical Physics, Moscow State University, 119991 Moscow, Russia}%

\author{\firstname{A.~I.}~\surname{Studenikin}}
\email{studenik@srd.sinp.msu.ru}
\affiliation{Department of Theoretical Physics, Moscow State University, 119991 Moscow, Russia}%

\begin{abstract}
Neutrino spin-flavour oscillations in the magnetic field, matter, and majoron fields are studied. The effective Lagrangian of Majorana neutrino interaction with a majoron field mediated by heavy neutrino is calculated. It is shown that the presence of a majoron field with a high energy density significantly shifts the resonances in neutrino oscillations, or even induces new resonances. The influence of the majoron suppresses oscillations for neutrino energies below 10 MeV. Additionally, the dependence of the oscillations probabilities on the Dirac and Majorana CP-violating phases becomes weaker under the influence of a dense majoron field. These findings can be applied to experiments that study neutrinos from supernovae, such as JUNO, DUNE, and the upcoming Hyper-Kamiokande experiment.
\end{abstract}

\maketitle

\section{Introduction}
The question of whether neutrinos are Dirac or Majorana particles is open. In the case of Majorana neutrino nature the problem of neutrino mass generation could be solved with seesaw mechanism. Among the many ways to introduce the seesaw mechanism, one stands out based on the decay of a heavy neutrino into a light one and an axion-like particle called majoron. This approach can guarantee a suitable ratio of the heavy and light neutrinos masses. The seesaw mechanism by majorons is given in \cite{Reig:2019}. 

Majorons can be born in a supernova core in the neutrino coalescence process ($\nu + \nu \rightarrow a$) \cite{Raffelt:2023} and participate in a process of supernova cooling. Moreover, due to their boson nature, majoron particles can form hypothetical astrophysical objects such as boson stars. And, of course, being one of the axion-like particles (ALPs), majoron is a popular candidate for the role of dark matter.

In this work, we continue our investigation of neutrino oscillations in astrophysical environments, focusing on the case when neutrinos are Majorana particle. In addition to studying neutrino oscillations in the presence of a magnetic field and matter, we also take into account the influence of majorons. In Section II, the effective majoron-neutrino interaction Lagrangian is derived, accounting for the neutrino interaction with a magnetic field and matter. Section III presents analytical results in the two flavours oscillations case, in which either a magnetic field or matter is absent. Finally, Section IV contains numerical results on the neutrino flavour and spin-flavour oscillations in the presence of the majoron and magnetic fields, and matter.

\section{Majorana neutrino interaciton with majoron, matter and magnetic fields in astrophysical environments}
\subsection{Majorana neutrino-majoron effective interaction}

Interaction between neutrino mass states  and majoron is described by the  Lagrangian \cite{Ak:2023}
\begin{equation}
    \mathcal{L}_a = -i\frac{a}{F}\bar n_i\left(g^{ij}_V(m_i - m_j) +\gamma_5g^{ij}_A(m_i + m_j)\right)n_j,
\end{equation}
where $n_i$ are the spinors ($i=1,2,3$) which consist of light neutrino mass state $\nu_i$ and corresponding heavy neutrino $N_i$ and defined as
\begin{equation}
    n_i = (\nu_i, N_i)^T.
\end{equation}
Consider a neutrino forward scattering process $a + \nu_i \rightarrow a +\nu_k$ intermediated by a heavy neutrino. Neutrino propagator is simplified to
\begin{equation}
    S_{c_{j}} = \frac{1}{k^\mu\gamma_\mu - M_j} \rightarrow -\frac{1}{M_j},
\end{equation}
where $M_j$ are masses of heavy neutrino. In this approximation the effective coupling constant has the following form
\begin{equation}\label{G}
    G^{ik} = \sum_j\frac{g_V^{ij}g_V^{jk}(m_i - M_j)(M_j - m_k) + g_A^{ij}g_A^{jk}(m_i + M_j)(M_j + m_k)}{F^2M_j}.
\end{equation}
Therefore, the effective Lagrangian of majoron-neutrino interaction after averaging over majoron background reads
\begin{equation}\label{Laik}
\mathcal{L}_{\left\langle a \right\rangle} = \frac{\rho_a}{m_a^2}G^{ik}\bar\nu_i\nu_k,
\end{equation}
where $\rho_a$ is an energy density of majoron,
\begin{equation}
\frac{\rho_a}{m_a^2} = \left\langle a^2 \right\rangle.
\end{equation}
To calculate (\ref{G}), it is necessary to take into account the dependence of $g_V$ and $g_A$ on the mixing of heavy and light neutrinos (see \cite{Ak:2023, Casas:2001}).
\begin{eqnarray}\label{gva}
	&g_V^{ij} = \frac{1}{2}\mathrm{Im }C_{ij},\\
	&g_A^{ij} = \frac{i}{2}\Big(\frac{1}{2}\delta_{ij} - \mathrm{Re }C_{ij}\Big),
\end{eqnarray}
where
\begin{eqnarray}\label{C}
	C =
	\begin{pmatrix}
		1 & i\sqrt{d_l}R^T\sqrt{d_h^{-1}} \\
		-i\sqrt{d_h^{-1}}R^*\sqrt{d_l} & 0
	\end{pmatrix}
	,
\end{eqnarray}
$d_l= \mathrm{diag}(m_1, m_2, m_3)$, $d_h= \mathrm{diag}(M_1, M_2, M_3)$ and $R = (R^T)^{-1}$ is a complex orthogonal matrix. Substitution (\ref{gva}) in (\ref{G}) yields
\begin{equation}\label{G_fn}
	G^{ik} = \frac{m_i}{4F^2}\delta_{ik}.
\end{equation}
Therefore, the effective Lagrangian (\ref{Laik}) reduces to
\begin{equation}\label{La}
\mathcal{L}_{\left\langle a \right\rangle} = \frac{\rho_a}{4F^2m_a^2}m_i\bar\nu_i\nu_i.
\end{equation}

It is obvious that Lagrangian (\ref{La}) shifts the neutrino mass, as will be seen below.

\subsection{Majorana neutrino interaction with a magnetic field and matter}

Majorana neutrino interaction with a magnetic field $\bm{B}$ is described (see \cite{Popov:2021,Popov:2023wif}) by the following Lagrangian
\begin{equation}\label{mag_field_int}
\mathcal{L}_{mag} = -\sum_{ik}\mu_{ik}\left[ \overline{(\nu_i^L)^c} \bm{\Sigma}\bm{B} \nu_k^L + \overline{\nu_i^L} \bm{\Sigma}\bm{B} (\nu_k^L)^c \right] =
-\sum_{\alpha\beta}\left[\mu_{\alpha \beta}^{(f)} \overline{(\nu_{\alpha}^L)^c} \bm{\Sigma}\bm{B} \nu_{\beta}^L - (\mu^{(f)})^\dag_{\alpha\beta} \overline{\nu_{\alpha}^L} \bm{\Sigma}\bm{B} (\nu_{\beta}^L)^c \right],
\end{equation}
where $\mu$ and $\mu^{(f)}$ are the neutrino magnetic moments matrix in the massive states basis and in the flavour states basis correspondingly. Note that in \cite{Popov:2021,Popov:2023wif} the hermitian conjugate in the last term of Eq. (\ref{mag_field_int}) is missing due to a typo, but is present in the formulas used for actual numerical computations of the oscillations probabilities.

Due to CPT-invariance and hermiticity of the interaction Lagrangian, the magnetic moments matrix of Majorana neutrinos in the massive neutrino states basis is antisymmetric and imaginary, and can be parametrized as follows \cite{Giunti:2014ixa}

\begin{equation}
\mu=\begin{pmatrix}
0 & i\mu_{12} & i|\mu_{13}| \\
-i|\mu_{12}| & 0 & i|\mu_{23}| \\
-i|\mu_{13}| & -i|\mu_{23}| & 0
\end{pmatrix}.
\end{equation}

Although in our numerical computations below we use the neutrino evolution equation in the mass basis, it is useful to consider the magnetic moments matrix in the flavour basis for illustrative purposes. The transition to the flavour basis in the case of Majorana neutrinos is performed by the following transformation

\begin{equation}\label{mm_flavour}
\mu^{(f)}=U\mu U^T,
\end{equation}
where $U$ is the neutrino mixing matrix. For the case of Majorana neutrinos the mixing matrix $U$ is given by

\begin{equation}\label{PMNS}
	U=\begin{pmatrix}
	1 & 0 & 0 \\
	0 & c_{23} & s_{23} \\
	0 & -s_{23} & c_{23}
	\end{pmatrix}
	\begin{pmatrix}
	c_{13} & 0 & s_{13} e^{-i\delta} \\
	0 & 1 & 0 \\
	-s_{13} e^{i\delta} & 0 & c_{13}
	\end{pmatrix}
	\begin{pmatrix}
	c_{12} & s_{12} & 0 \\
	-s_{12} & c_{12} & 0 \\
	0 & 0 & 1
	\end{pmatrix}
	\begin{pmatrix}
	e^{i\alpha_1} & 0 & 0 \\
	0 & e^{i\alpha_2} & 0 \\
	0 & 0 & 1
	\end{pmatrix}
\end{equation}
and contains three CP-violating phases: the Dirac CP-violating phase $\delta$ and the Majorana CP-violating phases $\alpha_1$ and $\alpha_2$.

Note that the transformation (\ref{mm_flavour}) differs for the one for the case of Dirac neutrinos that is given by $\mu^{(f)}=U\mu U^\dag$. This is due to the terms $\overline{(\nu_{\alpha}^L)^c} = (\nu_{\alpha}^L)^T C$ in the Lagrangian (\ref{mag_field_int}) that contain matrix transpose $U^T$ rather than hermitian transpose $U^\dag$. This difference leads to the new effects in neutrino oscillations in a magnetic field when the CP-violating phases are present in the mixing matrix \cite{Popov:2021,Popov:2023wif}. Below we also study these effects when neutrino interaction with majorons is taken into account.

In particular astrophysical environments like supernovae, dense matter is present. A Majorana neutrino interaction with matter is described by the Lagrangian

\begin{equation}\label{matter_int_majorana}
\mathcal{L}_{mat} = \sum_{\alpha} \frac{V^{(f)}_{\alpha}}{2} \left[\overline{\nu^L_{\alpha}} \gamma_0 \nu_{\alpha}^L - \overline{(\nu^L_{\alpha})^c} \gamma_0 (\nu_{\alpha}^L)^c  \right],
\end{equation}
where 
\begin{equation}
    V^{(f)} = \text{diag}\left(\frac{G_Fn_e}{\sqrt{2}} - \frac{G_Fn_n}{2\sqrt{2}}, - \frac{G_Fn_n}{2\sqrt{2}}, - \frac{G_Fn_n}{2\sqrt{2}}\right)
\end{equation}
is the Wolfenstein potential. Unlike for the case of Dirac neutrinos, right-handed fields $\nu_\alpha^R = (\nu^L_{\alpha})^c$ are not sterile since they correspond to antineutrinos in the Majorana theory.

\subsection{Neutrino evolution equation}

Using Eqs. (\ref{La}), (\ref{mag_field_int}) and (\ref{matter_int_majorana}), the following system of evolution equations for neutrinos interacting with majoron and magnetic fields, and matter can be obtained
\begin{equation}\label{dirac_eqn}
    \left( \gamma_\mu p^\mu - m_i - \frac{\rho_a}{4F^2m_a^2}m_i - V^{(m)}_{ii}\gamma_0\gamma_5\right)\nu_i(p) - \sum_{k\neq i}\left(-\mu_{ik}\bm{\Sigma}\bm{B} + V^{(m)}_{ik}\gamma_0\gamma_5\right)\nu_k(p)= 0,
\end{equation}
where $V^{(m)} = U^\dagger V^{(f)}U$ is the matter potential in the mass basis and $p^\mu$ is the neutrino 4-momentum. Since matter potential in the flavour basis $V^{(f)}$ is diagonal, it follows from Eq. (\ref{PMNS}) that $V^{(m)}$ does not depend on the Majorana CP-violating phases.

It is possible to rewrite Eq. (\ref{dirac_eqn}) in the Hamiltonian form
\begin{eqnarray}\label{equation_hamiltonian}
i\frac{\partial}{\partial t} \nu(p) = 
\begin{pmatrix}
H_{11} & H_{12} & H_{13} \\
H_{21} & H_{22} & H_{23} \\
H_{31} & H_{32} & H_{33}
\end{pmatrix} \nu(p),
\end{eqnarray}
where $\nu = (\nu_1, \nu_{2}, \nu_{3})^T$ and
\begin{equation}
H_{ik} = \delta_{ik} \gamma_0 \bm{\gamma}\bm{p} + \left(1+\frac{\rho_a}{4F^2m_a^2}\right)m_i\delta_{ik} \gamma_0 + \mu_{ik}\gamma_0\bm{\Sigma}\bm{B} + V_{ik}^{(m)}\gamma_5.
\end{equation}
Using (\ref{equation_hamiltonian}), it is possible to derive an expression for the probability of neutrino spin-flavour oscillations. Applying the approach developed in \cite{Popov:2021}, it is possible to obtain the probabilities of neutrino spin-flavour oscillations

\begin{equation}\label{prob}
P(\nu_\alpha^s \rightarrow \nu_\beta^{s'};x) = \Big| \sum_n \sum_{i,k} U_{\beta k}^{s'} U_{\alpha i}^s C_{nki}^{s s'} e^{-i E_n x} \Big|^2,
\end{equation}
where $\alpha,\beta = e,\mu,\tau$ are neutrino flavours, $E_n$ are the eigenvalues of the Hamiltonian (\ref{equation_hamiltonian})
and $U^s$ are the mixing matrices for left-handed  ($s=L$) and right-handed ($s=R$) neutrinos. In the considered case of Majorana neutrinos $\nu_i^R = (\nu_i^L)^c$, and then $U^L = U$, $U^R = U^*$.

The coefficient $C_{nki}^{ss'}$ has the following form:

\begin{equation}
C_{nki}^{ss'} = \braket{\psi_k^{s'}(0)|P_n|\psi_i^s(0)},
\end{equation}
where $P_n = \ket{n}\bra{n}$ are the projection operators, $\ket{n}$ are the eigenvectors of the Hamiltonian (\ref{equation_hamiltonian}), and $\psi_i^s(0)$ are the wave functions of the initial neutrino mass state $i$ with helicity s.

Finally, for the neutrino oscillations probability we get 
\begin{widetext}
\begin{equation}\label{prob_real}
P(\nu_{\alpha}^s \rightarrow \nu_{\beta}^{s'};x) = \delta_{\alpha \beta}\delta_{s s'} - 4 \sum_{n>m} \operatorname{Re}(A_{\alpha\beta nm}^{s s'}) \sin^2\left(\frac{\pi x}{ L^{osc}_{nm} }\right)
+ 2 \sum_{n>m} \operatorname{Im}(A_{\alpha\beta nm}^{s s'})  \sin\left(\frac{2\pi x} {L^{osc}_{nm}}\right),
\end{equation}
\end{widetext}
where
\begin{equation}
A_{\alpha\beta nm}^{s s'} = \sum_{i,j,k,l} (U_{\beta k}^{s'})^* U_{\alpha i}^s (U_{\beta l}^{s'})^* U_{\alpha j}^s \big(C_{nki}^{ss'}\big)^*C_{mlj}^{s s'}
\end{equation}
and
\begin{equation}\label{L}
L^{osc}_{nm} = 2\pi/ (E_n - E_m).
\end{equation}

In order to find exact analytical expression for $C_{nki}^{ss'}$, three particular cases are considered below.

\section{Analytical results}
First, we consider neutrino oscillations in the presence of electromagnetic and majoron fields in the two flavour approximation. It can be shown that in the two flavour case the mixing matrix of Majorana neutrinos contains one Majorana CP-violating phase (see, for example, \cite{Giunti:2010ec}). The mixing matrix can be written as
\begin{equation}
    U_{2f}^M =\begin{pmatrix}
        \cos\theta & \sin\theta \\
        -\sin\theta & \cos\theta
    \end{pmatrix} \cdot
    \begin{pmatrix}
        e^{i\alpha} & 0\\
        0 & 1
    \end{pmatrix},
\end{equation}
where $\alpha$ is the CP-violating phase.

In this case the oscillations probabilities are given by
\begin{eqnarray}\label{p1}
   P(\nu_e \rightarrow \nu_\mu;x) &=& \frac{\omega_{a}^2}{\omega_{a}^2 + \omega_B^2}\sin^22\theta\sin^2\left(\sqrt{\omega_B^2 + \omega^2_a} x\right), \\
   \label{p2}
    P(\nu_e \rightarrow \bar{\nu}_\mu;x) &=& \frac{\omega_B^2}{\omega_{a}^2 + \omega_B^2}\sin^2\left(\sqrt{\omega_B^2 + \omega^2_a}x\right), \\
    \label{p3}
    P(\nu_e \rightarrow \bar{\nu}_e;x) &=& 0,
\end{eqnarray}
where $\omega_B$ and $\omega_a$ are the magnetic and majoron frequencies respectively
\begin{eqnarray}
    &\omega_a = \left(1+\frac{\rho_a}{4F^2m_a^2}\right)\frac{\Delta m^2}{4E_\nu}, \\
    &\omega_B = \mu B_\perp \cos\alpha,
\end{eqnarray}
where $\Delta m^2$ is a neutrino mass square difference, $\mu = \mu_{12}=\mu_{21}$ is the transition (non-diagonal) magnetic moment and $E_\nu$ is a neutrino energy. 

Above a certain threshold energy $E_{thr}$ the magnetic frequency $\omega_B$ in (\ref{p1}), (\ref{p2}) and (\ref{p3}) is greater than $\omega_a$, and the flavour transitions $\nu_e \to \nu_\mu$ become suppressed. The higher the majoron density, the greater this threshold energy is.

In the case of absence of interaction with majorons and absence of CP-violation ($\alpha=0$), Eqs. (\ref{p1}) and (\ref{p2}) are similar to the expressions for the oscillations probabilities obtained in \cite{Kouzakov:2017}.

The second analytically solvable case describes neutrino oscillations in the presence of matter and majoron field. In the case of uniform matter the expression for the flavour transition probability has the following form
\begin{equation}
    P(\nu_e \rightarrow \nu_\mu;x) = \sin^2(\omega_{mat}x)\sin^22\tilde{\theta},
\end{equation}
where $\omega_{mat}$ is a matter frequency
\begin{equation}
    \omega_{mat} = \sqrt{\omega_a^2\sin^22\theta+\left(\omega_a\cos 2\theta - \frac{G_F}{\sqrt{2}}n_e\right)^2},
\end{equation}
and $\tilde{\theta}$ is an effective mixing angle
\begin{equation}
    \sin^2 2\tilde{\theta} = \frac{\omega_a^2\sin^22\theta}{\left(\omega_a\cos 2\theta - \frac{G_F}{\sqrt{2}}n_e\right)^2+\omega_a^2\sin^22\theta}
\end{equation}
As can be seen from the formula, the presence of majoron fields can shift the resonance of the oscillations. The resonant density for a fixed neutrino energy can be written as

\begin{equation}
    G_F n_e^{res} = \frac{\sqrt{2}\Delta m^2 \cos2\theta}{E_\nu} \left(1+\frac{\rho_a}{4F^2m_a^2}\right).
\end{equation}

Thus, the resonant density in the presence of a majoron field is shifted by the factor $1+\frac{\rho_a}{4F^2m_a^2}$ compared to the case of usual vacuum neutrino oscillations. The widths of the resonance is given by $\Gamma = \omega_a \sin2\theta$ and is also modified by the same factor $1+\frac{\rho_a}{4F^2m_a^2}$ compared to the vacuum case.


Finally, we consider resonant spin-flavour conversion $\nu_e \to \bar{\nu}_\mu$ of Majorana neutrinos in magnetic, majoron fields and matter accounting only for transitions between $\nu_e$ and $\bar{\nu}_\mu$. The oscillations probability is
\begin{equation}\label{RSF}
P(\nu_e \to \bar{\nu}_\mu;x) = \sin^22\bar{\theta} \sin^2\bar{\omega} x,
\end{equation}
where the effective mixing angle is given by
\begin{equation}\label{RSF_ampl}
\sin^2 2\bar{\theta} = \frac{\omega_B^2}{\omega_B^2 + \left[ \sqrt{2}G_F n_B (Y_e -1/2) - \omega_a/2 \right]^2},
\end{equation}
and the oscillations frequency is
\begin{equation}
\bar{\omega} = \sqrt{(\omega_B)^2 + \left[ \sqrt{2}G_F n_B (Y_e -1/2) - \omega_a/2 \right]^2}.
\end{equation}

Here $n_B = n_p + n_n$ is the baryon number density, $n_n$ and $n_p$ are the neutron and proton number densities correspondingly, and $Y_e = n_e/n_B$ is the electron fraction. Assuming that the supernova media under consideration is electrically neutral, we can write $n_e = n_p$.

In the case of absence of interaction with a majoron fields and Majorana CP-violating phase $\alpha$, Eq. (\ref{RSF}) is similar to the expressions for the resonant conversion from \cite{Akhmedov:1988uk, Lim:1987tk}.

As it follows form (\ref{RSF_ampl}), neutrinos undergo resonant conversion $\nu_e \to \bar{\nu}_\mu$ when the resonant condition

\begin{equation}
    Y_e^{res}= \frac{1}{2} + \frac{\Delta m^2 \left(1+\frac{\rho_a}{4F^2m_a^2}\right)}{4\sqrt{2}G_F n_B E_\nu}
\end{equation}
is satisfied. Provided that the interaction with the majoron fields is absent, for supernova neutrinos, that have characteristic energies $E_\nu \sim 10$ MeV, the resonant condition simply yields $Y_e \approx \frac{1}{2}$. majoron fields of high density shifts the resonance to higher values of the electron fraction $Y_e$. Since most of the supernova models predict electron fractions $Y_e$ that are not significantly higher that 0.5 \cite{Fischer:2009af}, this effect may lead to complete disappearance of the resonance. The widths of the resonance $\Gamma = \omega_B$ is unaffected by the interaction with majorons.

\section{Numerical results}
This section presents numerical results for the Majorana neutrino oscillations probability (\ref{prob_real}) for the case of the three neutrino flavours accounting for the interactions with matter, magnetic and majoron fields. The figures below demonstrate the dependence of the oscillations amplitudes on the electron number density, electron fraction, neutrino energy, and Dirac and Majorana CP-violating phases. The values of the neutrino mixing parameters we use for numerical computations are given in Table \ref{tab:table1}.

\begin{table}[h]
\caption{\label{tab:table1}%
Neutrino oscillations parameters according to \cite{Esteban:2024}.}
\begin{ruledtabular}
	\begin{tabular}{c|c|c|c|c|c}
		\textrm{Parameter}&
		\textrm{$\sin^2 \theta_{12}$}&
		\textrm{$\sin^2 \theta_{23}$}&
		\textrm{$\sin^2 \theta_{13}$}&
		\textrm{$\Delta m^2_{12}  / \text{eV} ^2$}&
		\textrm{$\left| \Delta m^2_{13} \right|$/eV$^2$}\\
		\colrule
		Value & 0.307 & 0.561 & 0.022 & 7.49$\times 10^{-5}$ & 2.534$\times 10^{-3}$
	\end{tabular}
\end{ruledtabular}
\end{table}

As measured by the GEMMA reactor neutrino experiment \cite{Beda:2012zz}, the effective neutrino magnetic moment upper bound is $\mu_{\nu} < 2.9 \times 10^{-11} \mu_{B}$. The Borexino and XENON collaborations \cite{Borexino:2017fbd,XENON:2022ltv} provide even stronger upper limits through the observation of solar neutrino fluxes. These limits are $\mu_{\nu} < 2.8 \times 10^{-11} \mu_{B}$ and $6.4 \times 10^{-12} \mu_{B}$ correspondingly. For our further analysis, we fix the values of the neutrino magnetic moments in the mass basis as follows: $|\mu_{12}| = |\mu_{13}| = |\mu_{23}| = 10^{-13} \mu_B$. Also, we set the strength of the magnetic field to $10^{13}$ Gauss.

To estimate the energy density of the majoron field, we use the value of non-thermal majoron production presented in \cite{Reig:2019}. For this case energy density has the following form
\begin{equation}
    \rho_a \sim v_\sigma^2m_a^2,
\end{equation}
where $v_\sigma$ is associated with vacuum expectation value of majoron field. To illustrate the influence of the majoron field on neutrino oscillations we use the upper limit of $v_\sigma = 10^{11}$ GeV. Also, we set coupling constant $F$ to $10^9$ GeV.
\begin{figure}[h]
\begin{minipage}[t]{.45\textwidth}
\centering
\includegraphics[width=\textwidth]{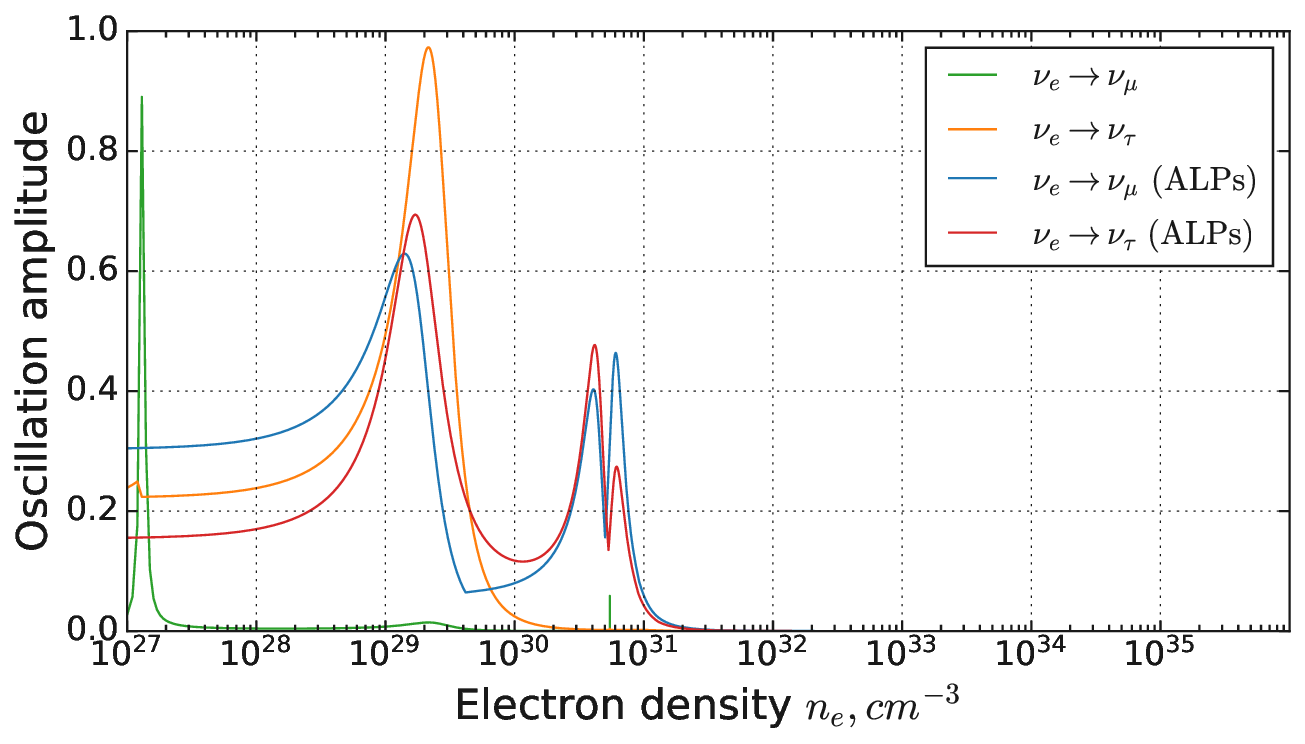}
\subcaption{ Neutrino-neutrino oscillations} 
\label{n_ea}
\end{minipage}
\hfill
\begin{minipage}[t]{.45\textwidth}
\centering
\includegraphics[width=\textwidth]{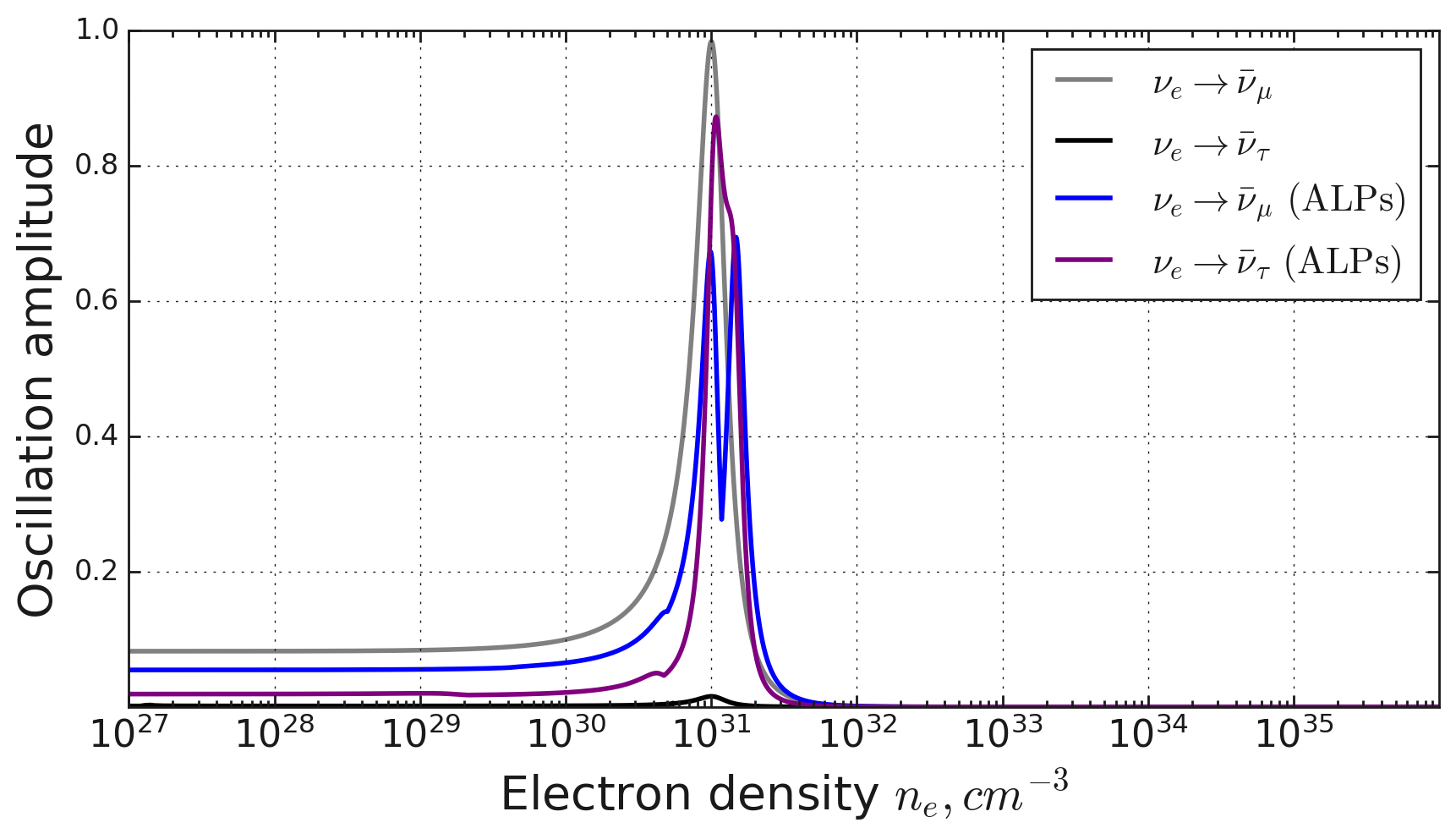}
\subcaption{Neutrino-antineutrino oscillations} 
\label{n_eb}
\end{minipage}
\caption{Dependence of the oscillation amplitudes on the electron number density. Dirac and Majorana CP-violating phases equal to zero, neutrino energy is set to be 10 MeV.}
\label{n_e}
\end{figure}

\begin{figure}[h]
\begin{minipage}[t]{.45\textwidth}
\centering
\includegraphics[width=\textwidth]{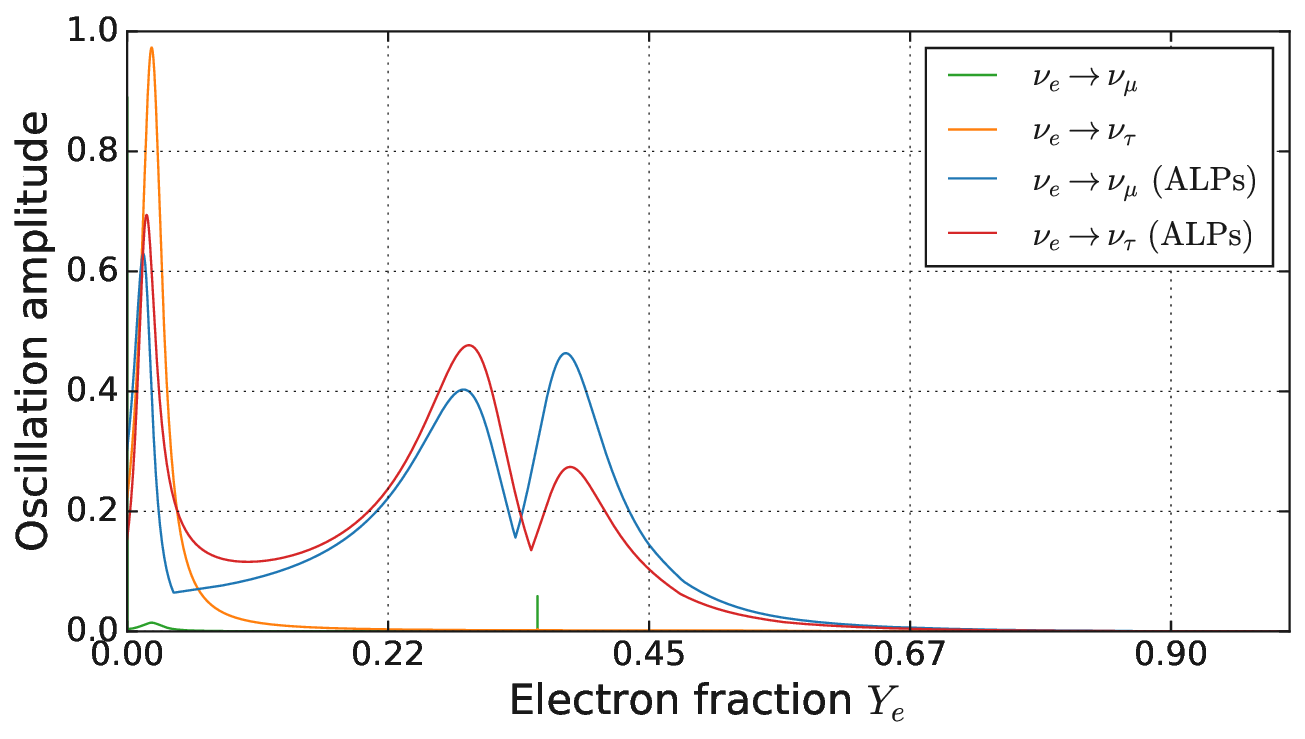}
\subcaption{Neutrino-neutrino oscillations.} 
\label{y_ea}
\end{minipage}
\hfill
\begin{minipage}[t]{.45\textwidth}
\centering
\includegraphics[width=\textwidth]{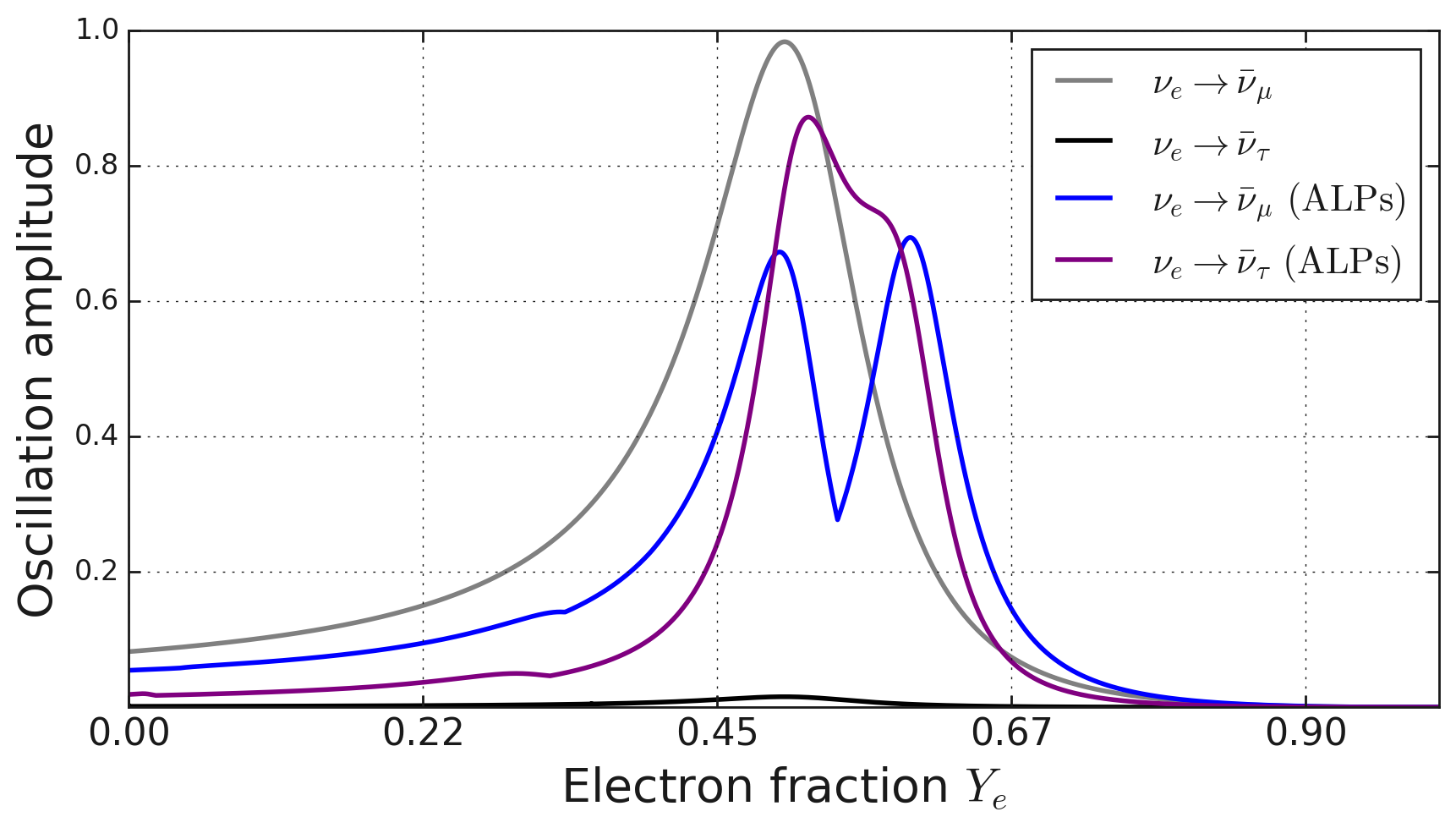}
\subcaption{Neutrino-antineutrino oscillations.} 
\label{y_eb}
\end{minipage}
\caption{Dependence of the oscillation amplitudes on the electron fraction. Dirac and Majorana phases equal to zero, neutrino energy is set to be 10 MeV.}
\label{y_e}
\end{figure}

Neutrino oscillations amplitudes as functions of electron number density $n_e$ and electron fraction $Y_e$ are shown in Fig. 1 and 2. In the case of neutrino-neutrino oscillations at Fig. \ref{n_e} and Fig. \ref{y_e}, majoron field shifts the positions of the resonance.  Moreover, two new resonances appear. In the case of neutrino-antineutrino oscillations, a new resonance appears. 

\begin{figure}[h]
\begin{minipage}[t]{.45\textwidth}
\centering
\includegraphics[width=\textwidth]{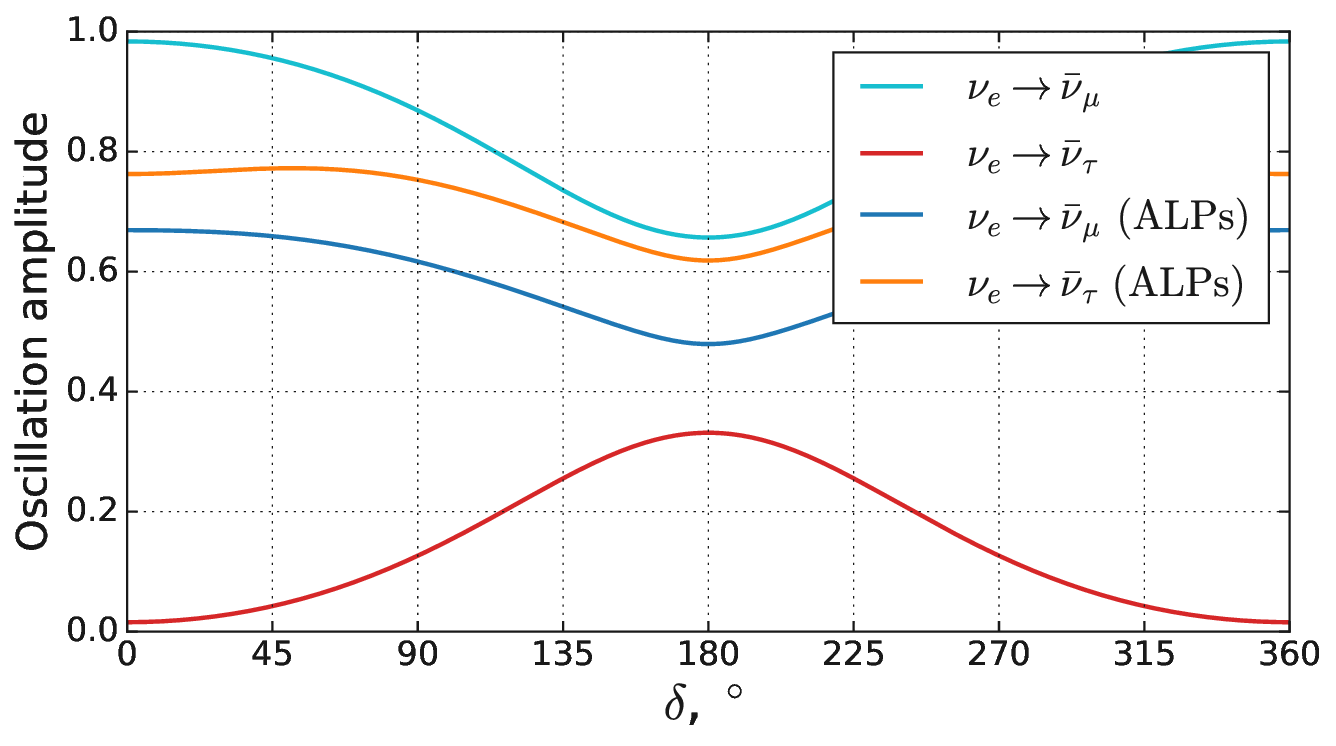}
\caption{Dependence of the neutrino-antineutrino oscillation amplitudes on the Dirac phase. Majorana phases equal to zero, neutrino energy is set to be 10 MeV, electron fraction equals to 0.5.} 
\label{delta}
\end{minipage}
\hfill
\begin{minipage}[t]{.45\textwidth}
\centering
\includegraphics[width=\textwidth]{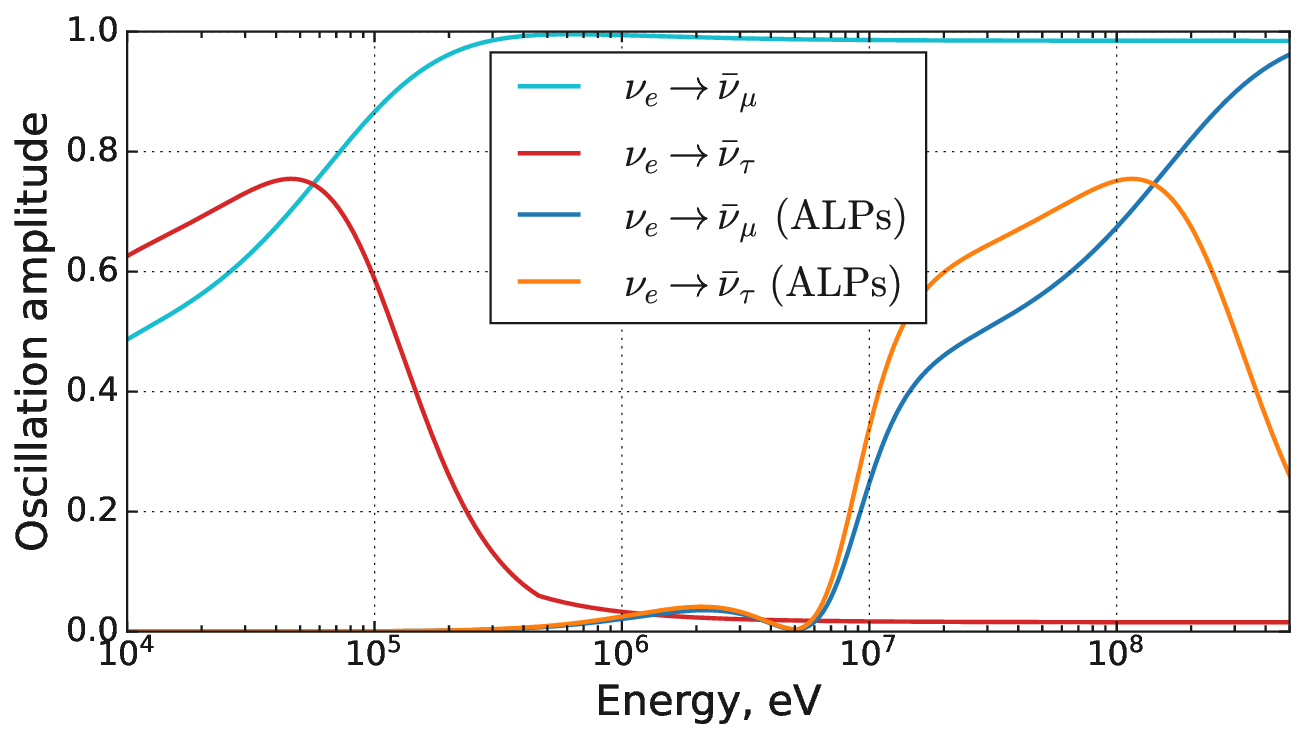}
\caption{Dependence of the neutrino-antineutrino oscillation amplitudes on the neutrino energy. Dirac and Majorana phases equal to zero, electron fraction equals to 0.5.} 
\label{energy}
\end{minipage}
\end{figure}

It can be seen from Fig. 3 that the presence of the majoron fields smears out the dependence of the neutrino oscillations amplitude on the Dirac CP-violating phase. The resonance of neutrino oscillations becomes slightly noticeable.


The dependence of the amplitudes of neutrino-antineutrino oscillations on the neutrino energy is shown in Fig. 4. The presence of the majoron field suppresses neutrino oscillations at low energies. The oscillations amplitude increases at the energies of 10 MeV and higher.

\begin{figure}[h]
\begin{minipage}[t]{.45\textwidth}
\centering
\includegraphics[width=\textwidth]{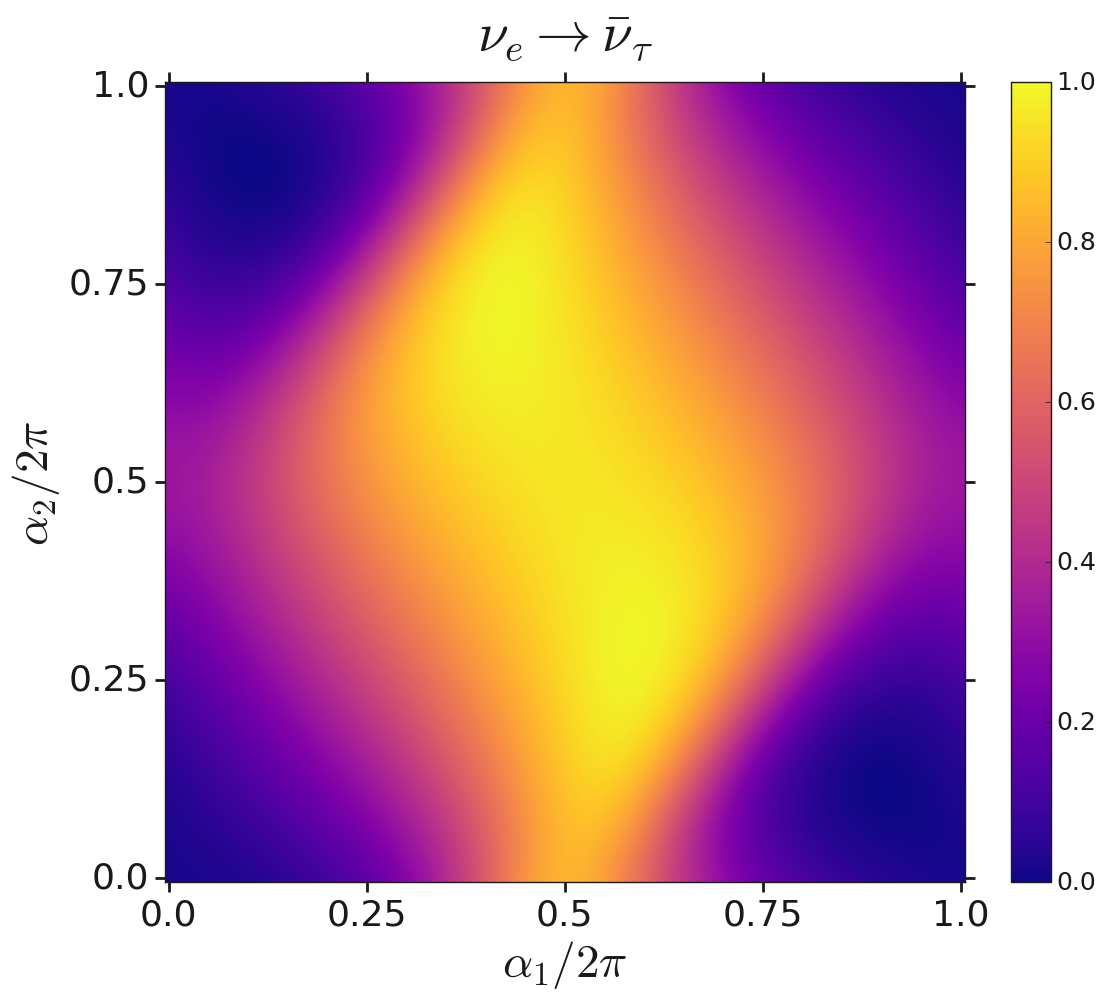}
\subcaption{Majoron field absence} 
\label{m_no_majoron}
\end{minipage}
\hfill
\begin{minipage}[t]{.45\textwidth}
\centering
\includegraphics[width=\textwidth]{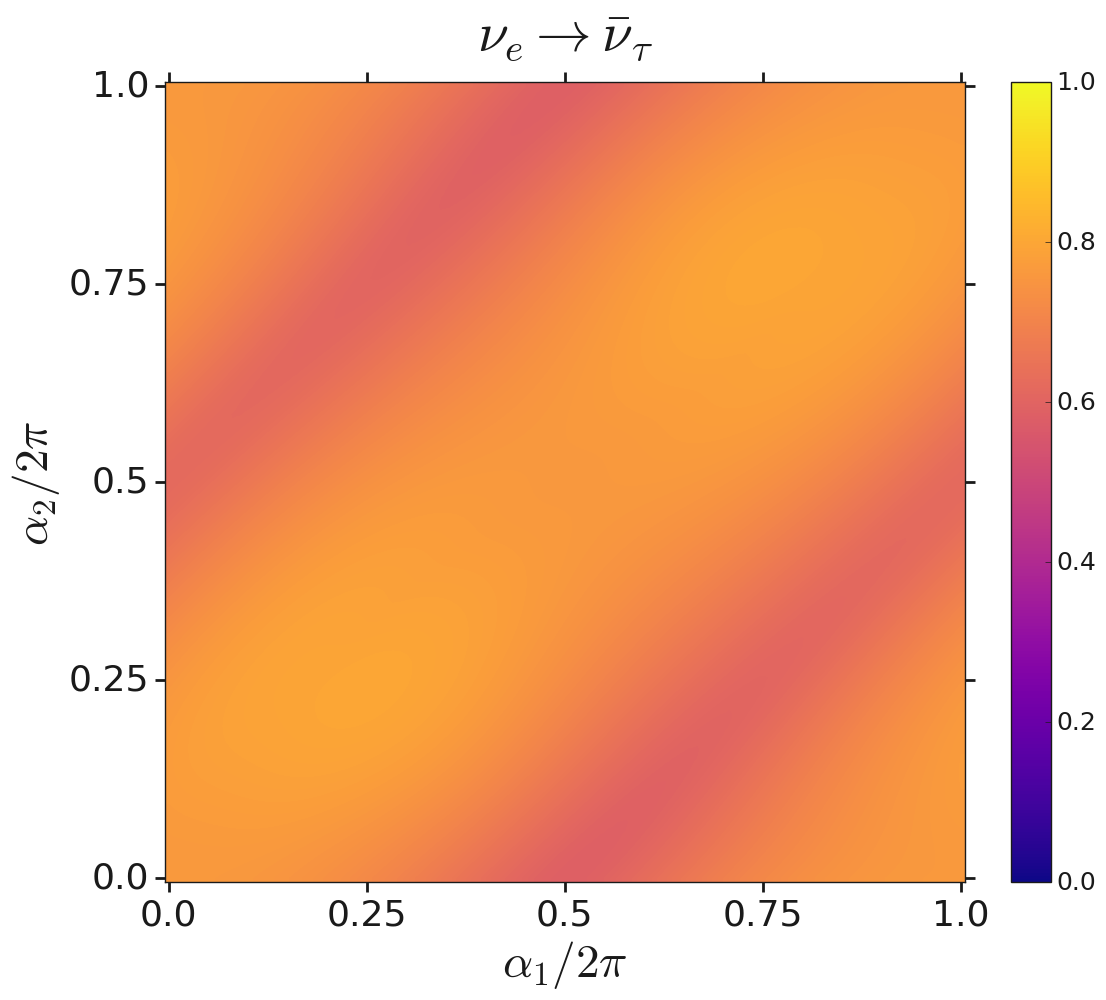}
\subcaption{Majoron field presence}
\label{m_majoron}
\end{minipage}
\caption{Dependence of the $\nu_e \to \bar{\nu}_\tau$ transition amplitude on the Majorana phases. Dirac phase equals to zero, electron fraction $Y_e$ equals to 0.5, neutrino energy is set to be 10 MeV.}
\end{figure}

The presence of majoron fields also washes out the dependence of neutrino oscillations probabilities on the Majorana CP-phases, as it can be seen from Fig. 5.


\section{Conclusions}
The comprehensive study of neutrino spin-flavour oscillations in the presence of magnetic, majoron fields and matter has revealed several important effects. Specifically, it has been demonstrated that the high-density majoron field not only shifts existing resonances, but also induces new resonance phenomena. 

An important observation is the suppression of neutrino oscillations for energies below 10 MeV under the influence of the majoron field. Furthermore, the analysis has shown a weakening of the dependence of the oscillations probabilities on both Dirac and Majorana phases in the presence of a dense majoron field. 

The findings of this research have direct implications for ongoing and future neutrino experiments, particularly those focused on supernova neutrinos. The JUNO, DUNE, and Hyper-Kamiokande experiments will benefit from these theoretical predictions, as they can incorporate the majoron field effects into their data analysis and interpretation. Future experimental observations may further validate these theoretical predictions and deepen our understanding of neutrino interactions in extreme conditions.

\begin{acknowledgments}
The work is supported by the Russian Science Foundation under grant No.24-12-00084. The work of A.L. has been supported by  the National Center for Physics and Mathematics (Project “Study of coherent elastic neutrino-atom and -nucleus scattering and neutrino electromagnetic properties using a high-intensity tritium neutrino source”).
\end{acknowledgments}

$\,$

$\,$

\appendix

\begin{thebibliography}{99}

\bibitem{Reig:2019} M.~Reig \textit{et al.}, JCAP \textbf{09} (2019) 029.
\bibitem{Raffelt:2023} D.~F.~G.~Fiorillo, G.~G.~Raffelt, E.~Vitagliano, Phys. Rev. Lett. \textbf{113} (2023), p. 021001.

\bibitem{Ak:2023} K.~Akita, M.~Niibo, J. High Energ. Phys. \textbf{2023} (2023), 132.
\bibitem{Casas:2001} J. A. Casas, A. Ibarra, Nucl.Phys. B. \textbf{618} (2001), 174.

\bibitem{Popov:2021} A.~Popov, A.~Studenikin, Phys. Rev. D \textbf{103} (2021), p. 115027.
\bibitem{Popov:2023wif}
A.~Popov, A.~Studenikin,
Phys. Part. Nucl. Lett. \textbf{21} (2024) no.3, 430-433.

\bibitem{Giunti:2014ixa}
C.~Giunti, A.~Studenikin,
Rev. Mod. Phys. \textbf{87} (2015), 531.

\bibitem{Giunti:2010ec}
C.~Giunti,
Phys. Lett. B \textbf{686} (2010), 41-43.

\bibitem{Kouzakov:2017} P.~Kurashvili \textit{et al.}, Phys. Rev. D \textbf{96} (2017), p. 103017.

\bibitem{Akhmedov:1988uk}
E.~K.~Akhmedov,
Phys. Lett. B \textbf{213} (1988), 64-68.

\bibitem{Lim:1987tk}
C.~S.~Lim and W.~J.~Marciano,
Phys. Rev. D \textbf{37} (1988), 1368-1373.

\bibitem{Fischer:2009af}
T.~Fischer \textit{et al.},
Astron. Astrophys. \textbf{517} (2010), A80.

\bibitem{Esteban:2024} I. Esteban, M.C. Gonzalez-Garcia, M. Maltoni \textit{et al.}, J. High Energ. Phys. \textbf{2024}, 216 (2025).
\bibitem{Beda:2012zz} A. Beda \textit{et al.} [GEMMA], { Adv. High Energy Phys.}  {\bf 2012}, 350150 ({2012}).
    
\bibitem{Borexino:2017fbd}
M.~Agostini \textit{et al.} [Borexino],
Phys. Rev. D \textbf{96} (2017) no.9, 091103.

\bibitem{XENON:2022ltv}
E.~Aprile \textit{et al.} [XENON],
Phys. Rev. Lett. \textbf{129} (2022) no.16, 161805,

\end{thebibliography}
\end{document}